\documentclass[12pt]{article}
\usepackage{cmb,epsfig}
\psfull

\newcommand{\ea}{{\it et~al.\/} }

\newcommand{\ho}{\mbox{$\mbox{H}_0$} }

\newcommand{\kmspmpc}{\,\hbox{km}\,\hbox{s}^{-1}\,\hbox{Mpc}^{-1}}

\newcommand{\eg}{{\it e.g.\ }}

\newcommand{\be}{\begin{equation}}
\newcommand{\ee}{\end{equation}}

\begin{document}
\heading{THEORETICAL IMPLICATIONS OF MICROWAVE BACKGROUND RADIATION ANISOTROPY
EXPERIMENTS}
\author{Graca Rocha and Stephen Hancock} {MRAO, Cambridge, UK.}{$\;$}

\begin{abstract}
 Observational results from several experiments like 
	COBE, SP, Saskatoon, PYTHON, ARGO, MAX, MSAM,
	Tenerife and CAT are considered and a comparison
	is made with predictions from several models.
	Conclusions are reached about the viability of current 
	structure formation models.	
\end{abstract}

\section{Introduction}
Cosmic microwave background (CMB) radiation anisotropies provide a
direct probe of the structure of the universe at early times and
constrain the values of key cosmological parameters such as $\Omega_0$, $H_0$ and $\Omega_{b}$.
Experiments are now producing reliable measurements of the anisotropy on a variety of angular scales. They
 provide useful information about the process of structure formation, avoiding the systematic uncertainties and biases of conventional techniques.
For Gaussian random fields all the statistical information is included in the angular power spectrum of the CMB temperature fluctuations. This being so, to determine the shape of this angular power spectrum constitutes one of the main goals of observational cosmology. 
We proceeded with a intercomparison of data and models in order to get the best-fit shapes of the angular power spectrum and constrain the model parameters.
We present an extension of the work carried out by Hancock \ea \cite{me96}, in which we used an analytic approximation to current CMB models and compared them with recent data.
Here we consider some of the exact COBE normalised angular power spectrum models for the four-year COBE data,
 and apply the analysis to 4 of the 8 binned angular power spectrum points \cite{tegmark} for the COBE experiment in conjunction with Tenerife, Python, South-Pole, Saskatoon, MAX, ARGO and CAT experiments.
We find that low values of $\Omega$ are excluded by these exact models, in agreement with the results presented by Hancock \ea \cite{me96}.
 For a flat standard Cold Dark Matter dominated Universe we use our
results in conjunction with Big Bang nucleosynthesis constraints to determine
the value of the Hubble constant as $\ho=30-70\kmspmpc$ for baryon fractions
$\Omega_b=0.05$ to $0.2$. For $\ho=50\kmspmpc$ we find the primordial spectral
index of the fluctuations to be $n=1.1 \pm 0.1$, in agreement with the
inflationary prediction of $n \simeq 1.0$. 
Flat models with a contribution from a cosmological constant $\Lambda$ seem to be consistent with the data.
 Models in which the fluctuations
originate from cosmic strings appear to be consistent with the observations if we do not consider the higher calibration Saskatoon data.
Textures model are only allowed for the lower calibration Saskatoon data.
A more extensive discussion of these models will appear in Rocha \ea \cite{gr} (in preparation).

\section{Models}

Models of structure formation \cite{be87,hu} predict the shape and amplitude of the CMB power spectrum and its Fourier equivalent, the autocorrelation function 
$C( \theta)=\langle \Delta T(n_{1}) \Delta T (n_{2}) \rangle$ where $n_{1}. n_{2}= \cos \theta$.
The intrinsic angular correlation function $C(\theta)$ may be expanded in terms of spherical harmonics:
$C(\theta)= \sum_{l \geq 2} ^{\infty} (2l+1) C_{l} P_{l}(\cos(\theta))/4 \pi$, where low order multipoles $l$ correspond to large angular scales $\theta$ and large $l$-modes to small angular scales.
The coefficients $C_{l}$ are predicted by the cosmological models and contain all of the relevant statistical information for models described by Gaussian random fields \cite{be87}.
We consider the following models:
flat CDM models with $\Omega_{b}$=0.01-0.3 and 
$H_{0}$=30-100 km s$^{-1}$ Mpc$^{-1}$ (G. Efstathiou, priv. comm.);
open models with values for ($\Omega_{0}$,$h_{0}$) (where $h_{0}=H_{0}/(100 \kmspmpc)$) of (0.1,0.75), (0.2,0.7), (0.3,0.65), 
(0.4,0.65), (0.5,0.6) (N. Sugiyama, priv. comm.);
models with a cosmological constant,
 $\Lambda$ and no reionization, with
 ($\Lambda$,$h_{0}$,$\Omega_{b}$) of (0.6,0.5,0.05), 
(0.6,0.5,0.03), (0.6,0.8,0.06),
 (0.7,0.5,0.05), (0.2,0.5,0.05);  a string model (J. Magueijo,  priv.comm.); and texture models with
 ($\Omega$,$h_{0}$,$\Omega_{b}$)=(1,0.5,0.05), (1,0.7,0.05) (N. Turok, priv. comm.).
\par
We also consider the case of tilted models with the spectral index $n$ of the primordial fluctuations varying in the range $n=0.7-1.4$ with ($h_{0}$,$\Omega_{b}$)=(0.3,0.2), (0.45,0.1), (0.5,0.07), for direct comparison with scale invariant flat CDM models.

\section{Data}

\begin{figure}
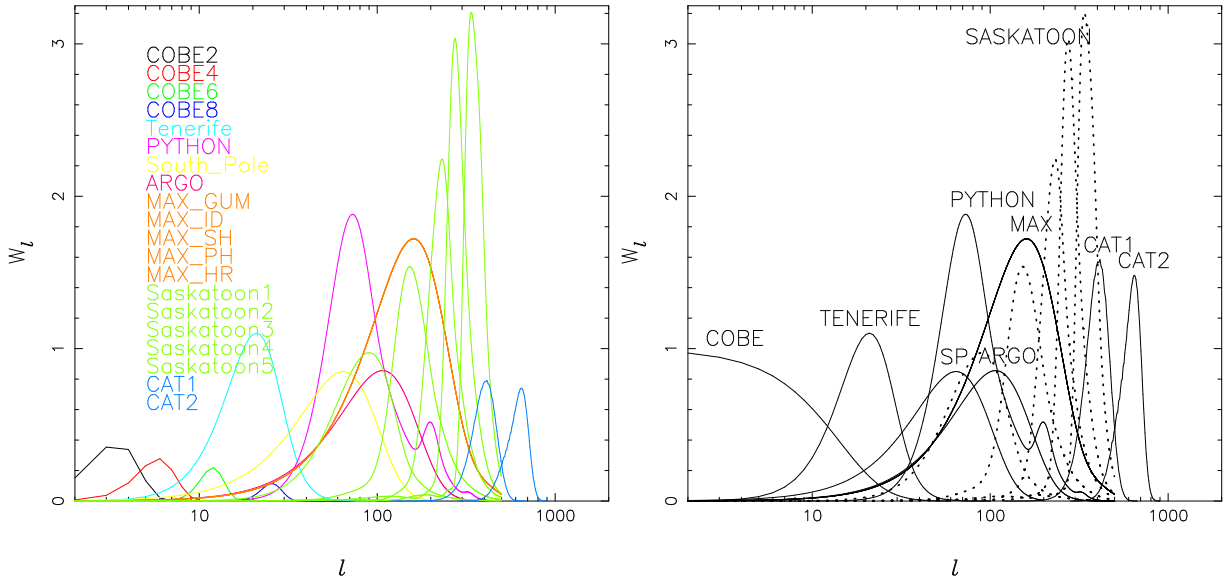

\hbox{%
\epsfig{file=wintegp.ps,width=3in,angle=270}

\epsfig{file=window.ps,width=3in,angle=270}

}
\caption{The window functions for the experiments listed in Table~1 \label{fig:windows1}}
\end{figure}

We consider all the latest CMB measurements, including new results from COBE, Tenerife, MAX, Saskatoon and CAT, with the exception of the MSAM  results and the MAX detection in the Mu Pegasi region ( for details on data selection see Hancock \ea \cite{me96}).
We also consider all 8 multipole bands of the angular power spectrum extracted from the 4-year COBE data \cite{tegmark}. 
In order to avoid eventual correlations between these data points we have applied the fitting analysis to the 4 even or 4 odd multipole bands in conjunction with the remaining data points.
The results are similar whether we use the even or the odd multipoles.
The use of 4 instead of one multipole is particularly useful in discriminating the value of the spectral index $n$.
The different experiments sample different angular scales according to their window functions \cite{whitewindow1,whitewindow2}, $W_{l}$, as shown in Fig.~\ref{fig:windows1}. On the left hand side we present the window functions for 4 of the 8 multipoles of the COBE experiment.
The conversion of data to a common framework in terms of flat bandpower estimates \cite{bond1, bond2} follows the same procedure as in Hancock \ea \cite{me96}.
In Table~1 we display these bandpower estimates $\Delta T_{l} \pm \sigma$ where $l_{l}$ and $l_{u}$ represent the lower and upper points at which the window function for each configuration reaches half of its peak value.
For each experiment the uncertainties of the intrinsic anisotropy level have been calculated using a likelihood analysis (see \eg Hancock \ea \cite{nature94}), which incorporates uncertainties due to random errors, theoretical uncertainty (cosmic variance) \cite{cosvariance2,cosvariance1} and enhancement of cosmic variance due to partial sky coverage (sample variance) \cite{samvariance}.
The errors in $\Delta T_{l}$ quoted in column 3 are 68 \% confidence limits and have been obtained by averaging the difference in the reported 68 \% upper and lower limits and the best fit $\Delta T_{l}$. This procedure introduces a small bias into the results since the likelihood function is in general only an approximation to a Gaussian distribution.

\section{Comparison}
\subsection{Method}

For a given theoretical model it is possible to compute the value of $\Delta T_{obs}$ one would expect to observe  using the angular power spectrum $C_{l}$'s as predicted by the model and the window function corresponding to the experimental configuration $W_{l}$. We convert this to the bandpower equivalent result $\Delta T_{l}$ and compare with $\Delta T_{l}^{obs}$. We then compute the chi-squared for this set of $C_{l}$'s
for the \rm{nd} data points in Table~1, and the relative likelihood function is formed according to: $L \propto exp(-\chi^{2} /2)$ ( for details see Hancock \ea \cite{me96}).
As mentioned above the process by which the error bars quoted in Table 1 are computed introduces a small bias into the results. In order to assess the significance of this we applied the fitting analysis to an alternative set of data which consisted of the mean value of the data and corresponding error bars. We find no significant difference in the results obtained. 
We also vary the power spectrum normalization within the 95\% limits for the COBE 4-year data \cite{bennett}.
The data included in the fit are those from Table 1, which, with the exception of Saskatoon, include uncertainties in the overall calibration.
There is a $\pm 14 \%$ calibration error in the Saskatoon data, but since the Saskatoon points are not independent this will apply equally to all five points \cite{sask}. The likelihood function is evaluated for three cases: (i) that the calibration is correct, (ii) the calibration is the lowest allowed value and (iii) the calibration is the maximum allowed value.
We consider that a given model offers an acceptable chi-squared fit when
 $P(\chi^{2}) \geq 0.05$.

\subsection{Results}

We have applied the analysis to the COBE normalised flat CDM models (provided by G. Efstathiou) considering only one multipole for COBE namely $C_{2}$ obtained from the value of the $Q_{rms-ps}$.
The models offering an acceptable chi-squared fit ($P(\chi^2) \ge 0.05$) to the CMB power spectrum whilst simultaneously satisfying nucleosynthesis constraints \cite{copi}, comprehend
$0.05\le \Omega_b \le 0.2$, $30\kmspmpc \le \ho \le 50 \kmspmpc$. 
Considering the highest Saskatoon data calibration the constraints become 
$0.1 \leq \Omega_{b} \leq 0.2$, $30\kmspmpc  \le \ho \le 35 \kmspmpc$. 
Allowing for the lowest Saskatoon data calibration relaxes the constraints up to $\ho=70
\kmspmpc$ and $0.02 \leq \Omega_{b} \leq 0.2$ (see Hancock \ea \cite{me96}). In general, recent optical and Sunyaev-Zel'dovich observations of
the Hubble constant \cite{ho3,ho1,ho2,lasenbyandhancock} imply \ho in the range
$50-80 \kmspmpc$.
  
Applying this analysis to the open models we conclude that $\Omega \leq 0.3$ is not compatible with the data, while the value of $\Omega =0.4$ is only allowed for the lower Saskatoon calibration data. A value of $\Omega =0.5$ offers an acceptable chi-squared fitting for both case (i) and (ii) calibrations.
This confirms the results obtained using the approximate formula of Hancock \ea reassuring us that the generalized parametrized formula  previously used constitutes a good approximation given the uncertainties in the data.

\begin{figure}
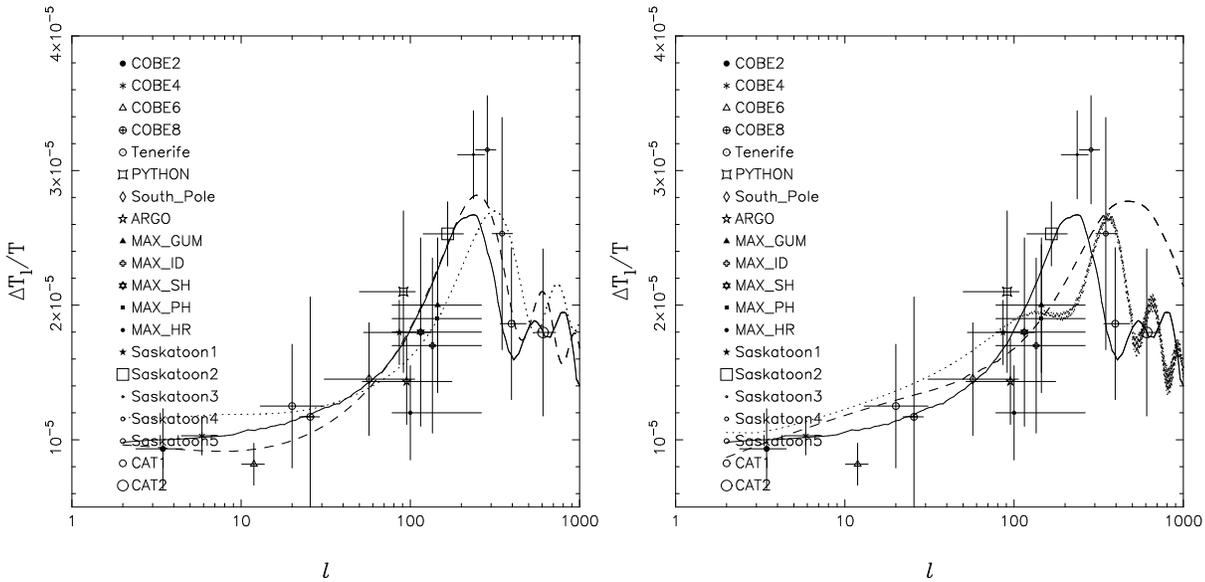

\hbox{%
\epsfig{file=cdm-open-lambd.1.ps,width=3in,angle=270}
\epsfig{file=tex-str-cdm.1.ps,width=3in,angle=270}
}
\caption{The data points from  table~1 compared to the exact forms of the $C_{l}$
for on the left hand side: an $\Omega=1$, $h_{0}=0.45$ and $\Omega_{b}=0.1$ standard CDM model (bold line),
an $\Omega=0.5$, $h_{0}=0.6$ open model (dot line) and a flat 
$\Omega_{\Lambda}=0.7$, $h_{0}=0.5$ and $\Omega_{b}=0.05$ model (dashed line).
For on the right hand side: the same standard CDM model compared with a cosmic string model (dashed line)
and an $\Omega=1$, $h_{0}=0.5$ and $\Omega_{b}=0.05$ texture model (dot line). 
\label{fig:models1}}
\end{figure}

The models with low $\Omega$ and a non-zero cosmological constant $\Lambda$ seem to be consistent with the data, with large values of $\Lambda$ providing the best fit.
In  Fig.~\ref{fig:models1} left hand side  we present the best candidates of both the open and $\Lambda$ models compared with a flat CDM model with $h_{0}=0.45$ and $\Omega_{b}=0.1$ assuming case (i) calibration. 
Models in which structure formation is initiated by cosmic strings \cite{strings} are consistent with the data, but are excluded for the higher calibration Saskatoon data. Texture models are also only allowed if one considers the lower calibration for the Saskatoon data.  
In  Fig .~\ref{fig:models1} right hand side we present a cosmic string model and the best candidate of the texture models (with $\Omega=1$, 
$h_{0}=0.5$ and $\Omega_{b}=0.05$) for comparison with the flat CDM model of Fig.~\ref{fig:models1} using case (i) calibration.

\begin{figure}
\hbox{%
\epsfig{file=tilt-50-07.1.ps,width=3in,angle=270}
\epsfig{file=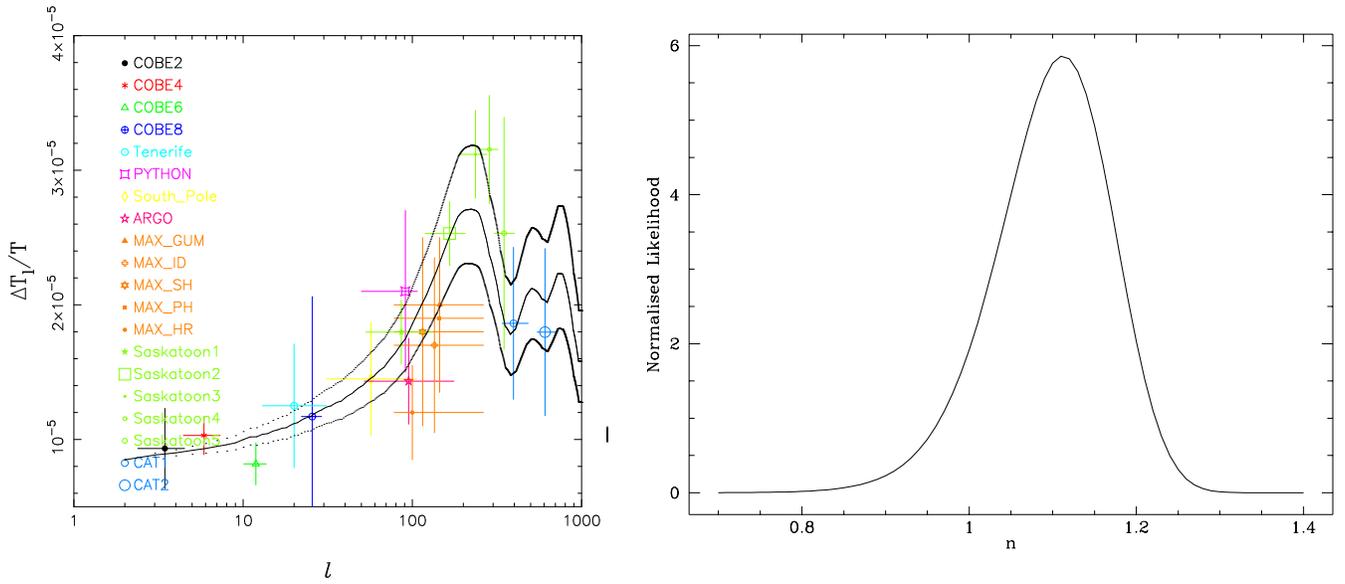,width=3in,angle=270}
}
\caption{ Tilted model with $h_{0}=0.5$, $\Omega_{b}=0.07$, $n=1.1$ and its 68 \%
confidence interval (left hand side), likelihood function for the parameter $n$ (right hand side). \label {fig:tilt1}} 
\end{figure}

We also consider some of the best candidates of the flat CDM models: (1) $h_{0}=0.3$, $\Omega_{b}=0.2$, (2) $h_{0}=0.45$, $\Omega_{b}=0.1$ and (3) $h_{0}=0.5$, $\Omega_{b}=0.07$ in which we allow for the tilting of the power spectrum to vary within $n=0.7-1.4$. 
The fitting for the spectral index $n$ of the primordial fluctuations was done considering two sets of data: (a) one multipole, (b) 4 multipole bands for the COBE experiment and case (i) for the calibration. For case (a) we get for model (1) a value of $n=0.94 \pm 0.06$, for model (2) $n=1.03 \pm 0.07$ and for model (3) $n=1.08 \pm 0.07$. For case (b) we find for model (1) $n= 0.99 \pm 0.07$, for model (2) 
$n=1.08 \pm 0.07$ and for model (3) $n=1.11 \pm 0.07$ (68 \% confidence interval).
Allowing for the three calibration cases and taking extreme limits the uncertainties are altered as follows: for case (a) for model (1) $n=0.94 \pm 0.1$, model (2) $n=1.03 \pm 0.1$ and model (3)
$n=1.08^{+0.09}_{-0.12}$ while for case (b) for model (1) $n=0.99 ^{+0.11}_{-0.13}$, model (2) $n=1.08 \pm 0.1$ and model (3) $n=1.11^{+0.09}_{-0.11}$.
 In particular fixing $H_{0}=50 \kmspmpc$ we find $n=1.1 \pm 0.1$ (68 \% confidence interval) allowing for the three calibration cases and taking extreme limits.
This tight limit rules out a significant contribution from gravity wave background for these models, in the case of power law inflation, but is consistent with the prediction of
 $n \simeq 1.0$ for scalar fluctuations generated by inflation. 
In  Fig.~\ref{fig:tilt1} left hand side  we present the model (3) and its 68 \% confidence interval assuming the case (i) for the calibration. On the right hand side is displayed the likelihood function for the parameter $n$ for model (3) which peaks at $n=1.1$ showing a significant estimate of the spectral index parameter based on the actual CMB data.

\begin{table*}[p]
\begin{center}
\caption{Details of data results used}
\begin{tabular}{|l|c|c|c|c|c|c|} \hline
Experiment & $\Delta T_l$ ($\mu$K) & $\sigma$ ($\mu$K) & $l_e$ & $l_l$  &$l_u$ & Ref
\\ \hline

 COBE2  &        25.4 &   8.1 & 3 & 3&  12& \cite{tegmark}    \\
 COBE4  &       28.1   & 3.9 & 5 & 5 & 8&\cite{tegmark} 	\\
 COBE6  &       22.3   & 4.2 & 11 & 10 & 15&\cite{tegmark} \\
 COBE8  &       31.9   & 24.3 & 25 & 22 & 30&\cite{tegmark} \\
 Tenerife &     34.1   & 12.5  &20 & 13 & 31&\cite{hancock}	\\
 PYTHON &       57.2 &   16.4&  91 & 50 & 107&\cite{python}\\
 South Pole &   39.5  &  11.4 & 57 & 31 & 106&\cite{spole}	\\
 ARGO        &  39.1   & 8.7 & 95 & 52 & 176&\cite{argo}	\\
 MAX GUM  &     54.5 &   13.6 & 145 & 78 & 263&	\cite{max}\\
 MAX ID       & 46.3   & 17.7 & 145 & 78 & 263&\cite{max}	\\
 MAX SH      & 49.1   & 19.1 & 145 & 78 & 263&\cite{max}	\\
 MAX PH      &  51.8   & 15.0 & 145 & 78 & 263&\cite{max}	\\
 MAX HR      &  32.7   & 9.5 & 145 & 78 & 263&\cite{max}	\\
 Saskatoon1 &   49.0   & 6.5 & 86 & 53 & 132 &\cite{sask}\\
 Saskatoon2   & 69.0   & 6.5  &166 & 119&  206&\cite{sask}\\
 Saskatoon3   &85.0    &8.9  &236  &190 & 274&	\cite{sask}\\
 Saskatoon4   & 86.0   & 11.0 & 285 & 243 & 320&\cite{sask}\\
 Saskatoon5   & 69.0   & 23.5 & 348 & 304 & 401&\cite{sask}\\
 CAT1         & 50.8   & 15.4 & 396 & 339 & 483&\cite{cat}	\\
 CAT2        &  49.0   & 16.9 & 608 & 546 & 722&\cite{cat}	\\
 
\hline
\end{tabular}
\end{center}
\end{table*}

%
\vfill

\end{document}